**Sagnac Electron Interference as a Probe of Electronic Structure**

Neda Lotfizadeh[1], Mitchell J. Senger[2], Daniel R. McCulley[2], Ethan D. Minot[2] and Vikram V. Deshpande[1*]

[1]Department of Physics and Astronomy, University of Utah, Salt Lake City, Utah, 84112, USA.

[2]Department of Physics, Oregon State University, Corvallis, OR, 97331, USA.

*Correspondence to:  vdesh@physics.utah.edu.

**Abstract**

Electronic analogues of optical interferences have been used to investigate quantum phenomena in condensed matter. However, the most sensitive, Sagnac optical interference, has not been realized in the electronic domain. We report electronic Sagnac interference using ultraclean carbon nanotubes. We show that Sagnac interference manifests as conductance oscillations versus gate voltage. Compared to Fabry-Perot conductance oscillations, Sagnac oscillations have a larger period and are more robust to temperature. Using nanotubes of known structure, we show that Sagnac oscillations are a probe of nanotube chirality in a transport measurement. Furthermore, upon tuning the system through a topological phase transition in a magnetic field, we find that the Sagnac interference undergoes a $\pi$ phase-shift. Our interferometry may be used for electronic structure characterization and study of many-body & photogalvanic effects in these model chiral one-dimensional systems.

**Introduction**

Quantum interferences of electron waves in devices with size close to the electron's coherence length have been instrumental in revealing various quantum and interaction effects (*1–6*).

Electronic analogues of Fabry-Perot (*7*) and Mach-Zehnder (*8*) interferometers have stimulated extensive research over the past two decades. Fabry-Perot (FP) interference (*7*) was experimentally discovered in single-wall carbon nanotubes (CNTs) as conductance oscillations as a function of a nearby gate voltage and has been the subject of numerous studies. Since its observation in GaAs–AlGaAs heterostructures, Mach-Zehnder interference (*8*) has been used in fields such as the fractional quantum Hall effect for detection of non-Abelian statistics. In comparison, common-path interferometers are known to be more robust to environmental perturbations. Among the most popular common-path interferometers is the one based on the Sagnac effect (*9*), occurring when waves of the same frequency travel at different velocities as they move in opposite directions along the same path. Traditionally this interference effect has been used to detect minute changes in the rotational velocity of optical ring interferometers and has been proposed as the basis for future-generation LIGO detectors. Loop-less versions of this interferometer have been realized using fiber optics with light of two orthogonal polarizations and applied toward ultra-sensitive magnetic detection (see e.g. ref *10*). However, Sagnac interference has not been realized in the electronic domain and is the main focus of the current work.

For electron waves in CNTs, Sagnac interference is predicted to occur without a ring interferometer and to be sensitive to asymmetries in electronic band structure (Fig. 1A and B) (*11, 12*). A forward-moving electron in one sub-band of the nanotube (the K sub-band) has a different phase velocity than a forward-moving electron in the other sub-band (the K′ sub-band). A band structure schematic for a metallic CNT is shown in Fig. 1B where the phase velocity is $E/\hbar k$ ($E$ is electron energy and $k$ is electron momentum along the axis of the CNT relative to the K or K′ point). A direct analogy can be made with Sagnac's original ring-shaped interferometer. Interference between forward/backward moving paths in the K sub-band of a nanotube ring (Fig.

1A) is equivalent to interference between K and K′ forward moving paths. This equivalence is a result of the time-reversal symmetries shown in Fig. 1B.

Previously, suppressed conductance at elevated temperatures in a nanotube loop was interpreted in terms of Sagnac interference (*11*), though the same effect has been observed due to substrate disorder. More recently, Dirnaichner et al. (*13*) obtained secondary conductance oscillations versus gate voltage in one nanotube device by averaging over Fabry-Perot oscillations. Both studies, however, did not have structural information of their nanotube to confirm their findings and had significant experimental uncertainty, making it impossible to extract information about the underlying electronic structure.

To eliminate disorder effects, our CNTs are grown across electrodes using chemical vapor deposition on prepatterned substrates (*14–19*) (see Supplementary Materials). Such devices have unveiled phenomena such as Wigner crystallization (*16*, *19*), spin-orbit coupling (*17*) and strong electro-mechanical coupling (*18*) previously hidden by disorder. These devices have all been in the sub-micron quantum-dot regime and mesoscopic transport in long ultraclean open systems with transparent contacts, without complications of charging effects e.g. Coulomb blockade, has not yet been explored. Our study addresses this deficit in the nanotube literature; furthermore, our longer nanotubes allow for optical characterization of CNT structure.

**Results**

Figure 1C shows a scanning electron microcopy image of a representative CNT. Our nanotubes are suspended over a ~2 μm trench, with two gates located at the bottom of the trench (Supplementary Materials Fig. S1). Fig 1D plots the conductance of a device D1 with measured length 2.2 μm as a function of gate voltage $V_g$ at T = 1.5 K. The regularity and stability of our data indicates that our devices are high-quality and defect-free. Due to the transparency of the contacts, the CNT acts as a waveguide at low temperatures. A small amount of scattering is expected at the interface between CNT and the metal contacts. Thus, interference between propagating and scattered electron waves with $\Delta\varphi = 2\pi n$ phase difference (where $n$ is an integer), can produce Fabry-Perot (FP) oscillations (7). We attribute the observed fast oscillations in conductance (period ~80 mV) to this FP mechanism. Interestingly, we observe a second periodicity of conductance oscillations with a much longer period (~2 V). We have observed this combination of short-period and long-period oscillations in over 30 nanotube devices. The nature of this long-period (slow) conductance oscillation is the central topic of our study.

Temperature dependent measurements of the conductance oscillations are shown in the inset of Fig. 1D. With increasing temperature, the rapid FP oscillations disappeared at ~5 K while slow oscillations survived up to 30 K. Slow oscillations survived as high as 60 K in other devices (see Supplementary Materials Fig. S2). Additional information was obtained from a variable length measurement (Fig. 2 for device D2). First, the Fermi energy was varied in the whole length of the CNT by operating both gate electrodes as a single gate (black data). Second, the Fermi energy was tuned in only one half of the CNT (red data). For the half-length measurement, opposite voltages were applied to the two gates, with one voltage held fixed while the other gate was varied. Thus, by using split gate voltages of opposite polarity, a high-transparency tunnel barrier was created in the center of CNT (20). The insets show differential conductance $dI/dV_{sd}$ as a function of $V_g$ and source-drain bias voltage $V_{sd}$ for the whole nanotube and half nanotube. The waveguide length can be obtained by setting the round-trip phase shift in FP interference (7), $2LeV_c/\hbar v_f$, equal to $2\pi$, where $V_c$ is the height of the rhombic pattern in Fig. 2 inset and $v_f$ is the Fermi velocity. Using $V_c$ = 1.3 mV (2.1mV) for the whole (half) CNT, we obtain appropriate values for the length (half-length) of the device. Comparing the black data (full length) and the red data (half-length), we note that both the FP oscillation period and the slow oscillation period change by the same factor of almost 2. This indicates that the same length scale is responsible for both interference effects.

Previous theoretical studies predicted that quantum interference effects beyond FP oscillations can arise in CNTs with open contacts (*11–13*, *21*, *22*). In particular, the theoretical picture of Refael and co-workers (*11*, *12*) explains our observed long-period oscillations and a series of associated consequences. We propose that the long-period oscillations are due to the small phase accumulation between two co-propagating electron beams that occupy different bands. Because this phase accumulation per unit energy is smaller than the phase accumulation in any one band, the effect should be less affected by thermal smearing of the Fermi level and should be observable at higher temperatures than FP interference (*11*, *12*). Our observed enhancement of the temperature scaling of Sagnac interference over Fabry-Perot, by a factor of 6-8, matches well with these theoretical predictions (*11*, *12*).

Further, Sagnac conductance oscillations should be sensitive to the chiral angle of the CNT. Detuning of the phase velocities in the K and K′ sub-bands (Fig. 1B) is caused by trigonal warping (*22*, *23*), i.e. trigonal distortion of the Dirac cones in the graphene bandstructure away from charge neutrality, due to the trigonal symmetry of graphene. The effect of trigonal warping on CNT band structure is very sensitive to chiral angle. Armchair CNTs (chiral angle = 0°) have maximal detuning between K forward/back (see Fig. 1B), while zig-zag CNTs (chiral angle = 30°) have no detuning. Thus, the period of the Sagnac oscillations should be a fingerprint of the CNT structure that can be detected in a transport measurement.

We examined the relationship between the long-period oscillation and nanotube chirality by performing experiments on three CNTs of known chiral index. Chiral index was determined using scanning photocurrent spectroscopy (see Supplementary Materials) to identify the energies of two or more exciton resonances in the as-fabricated suspended CNT device (*24*). A representative photocurrent spectrum is shown in Fig. 3A (see also Fig. S3). Chiral angles of 16.1°, 19.7° and 20.5° were determined for devices D3, D4 and D5 respectively. Using measured chiral angle, we calculated the difference between phase velocities of K and K' electrons at the Fermi energy using the tight-binding band structures (*25*). We overlaid the calculated phase difference between two interfering beams, $\Delta\varphi = (k_L - k_R)L$, as a function of gate voltage $V_g$ (using a conversion factor $\alpha$ which matched well with the experimentally determined gate efficiency) upon the measured conductance in D3, D4 and D5 as shown in Fig. 3B. Peaks and dips in the measured slow oscillations of conductance versus $V_g$ corresponded well with accumulated phase differences of 0 and π. To isolate the effect of chirality from device specific parameters (length $L$ and gate efficiency $\alpha$), we also plotted the slow oscillation period $\Delta V_g^s$ versus slow-oscillation index $n$ (Fig. 3C). We found good fits to the expression $\Delta V_g^s = \beta(\sqrt{n} - \sqrt{n-1})$, based on ref. (*21*), for D3, D4 and D5, where the fit parameter $\beta$ is expected to vary inversely with device parameters $\alpha$ and $L$. Plotting $\beta\alpha L$ versus chiral angle $\theta$ in Fig. 3C (inset), we found that the devices follow the expected $1/\sin 3\theta$ dependence (*22*). Thus, armed only with knowledge of device parameters viz. length and gate efficiency, one may use transport measurements to characterize the structure of nanotubes using the Sagnac interference effect.

Since our Sagnac oscillations arise from an electronic interference between two different sub-bands (or valleys), the oscillations may be sensitive to changes in the topology of the band-structure, for example upon application of a magnetic field. The topological winding number of the nanotube bands can change when an applied axial magnetic field crosses a critical value (*26*, *27*). When sweeping through this critical field, one of the 1D sub-bands of the nanotube crosses the 2D Dirac point of the underlying graphene band structure (Fig. 4D). Before and after this crossing, the 1d sub-bands (K and K′) differ in their winding number by unity or 1D topological invariant 'Zak phase' (*28*, *29*) by π.

Figure 4A shows the evolution of conductance of device D3 upon the application of a B-field parallel to the nanotube. The Sagnac interference undergoes a π phase shift around a critical field $B_{cr}$ ~ 4-6 T, while the FP oscillations do not change appreciably. Fig 4B shows line traces at 0 and 9 T clearly indicating the same. Concurrent with the π phase shift at $B_{cr}$, we also observe a minimum in the transport gap. Figure 4C shows the valence band edge of D3. With increasing B-field, the valence band edge approaches the conductance band edge below $B_{cr}$. Above $B_{cr}$, the valence band edge retreats from the conductance band.

The conversion of constructive Sagnac interference into destructive (and vice versa) upon the topological phase transition can arise if the pseudospins on the two valleys become orthogonal. Pseudospin is a degree of freedom associated with the pair of $2p_z$ orbitals in each unit cell of the graphene lattice (*30*, *31*). A flipping of pseudospin is expected for carriers at the top/bottom of the valance/conduction band when a 1d sub-band crosses the 2D Dirac point of the underlying graphene band structure (Fig. 4D). In our experiment, several factors affect the evolution of the pseudospin as a function of applied magnetic field. In non-interacting electron models, nominally metallic nanotubes such as the ones used in this study have a small transverse momentum relative to the center of the 2D Dirac point. This transverse momentum is caused by the curvature of the carbon lattice and is expected to be $\Delta k_\perp = \gamma \cos(3\theta)/d^2$ (where $\gamma$ = 0.0436 nm, $\theta$ is the chiral angle and $d$ is the diameter) (*32*). This transverse momentum can be compensated (*33*) with a small magnetic field $4\hbar\Delta k_\perp/de$ (~ 5 T for D3) at which point the topological transition occurs. Our interferometry thus directly detects this transition in the form of the π phase-shift.

**Discussion**

Electron-electron interaction effects add a complicating consideration to our observations. Even when the transverse momentum $\Delta k_\perp$ is compensated, residual gaps (up to half an eV) have been observed in ultra-clean nanotubes (*24*, *34*). The effect of these interaction-induced gaps on pseudospin has not yet been explored theoretically. The presence of a large gap at $B_{cr}$ could work to give rise to pseudospin flips at the comparatively small (< 100 meV) doping levels in our experiment. Our measurements are a unique probe of such effects, though further experimental and theoretical study of the interplay of topological transitions with many-body interactions is necessary, and of increasing interest in CNTs for reasons of realizing Majorana fermions in these model 1D systems.

In summary, our observation of the electronic Sagnac interference provides key electronic structure information in a transport measurement, which has remained hidden from all existing electrical techniques. Our chirality-sensitive effect also provides a new window to the consequences of inversion-symmetry-breaking through a study of some of the same photogalvanic effects (e.g. ref. *35*) predicted for non-centrosymmetric Weyl semimetals.

**Acknowledgments**

We thank Paul McEuen, Dmytro Pesin and Massimo Rontani for helpful discussions.

**Funding:** Work performed in Oregon was supported by the National Science Foundation under Grant No. 1709800. A portion of device fabrication was carried out in the University of California Santa Barbara (UCSB) nanofabrication facility.

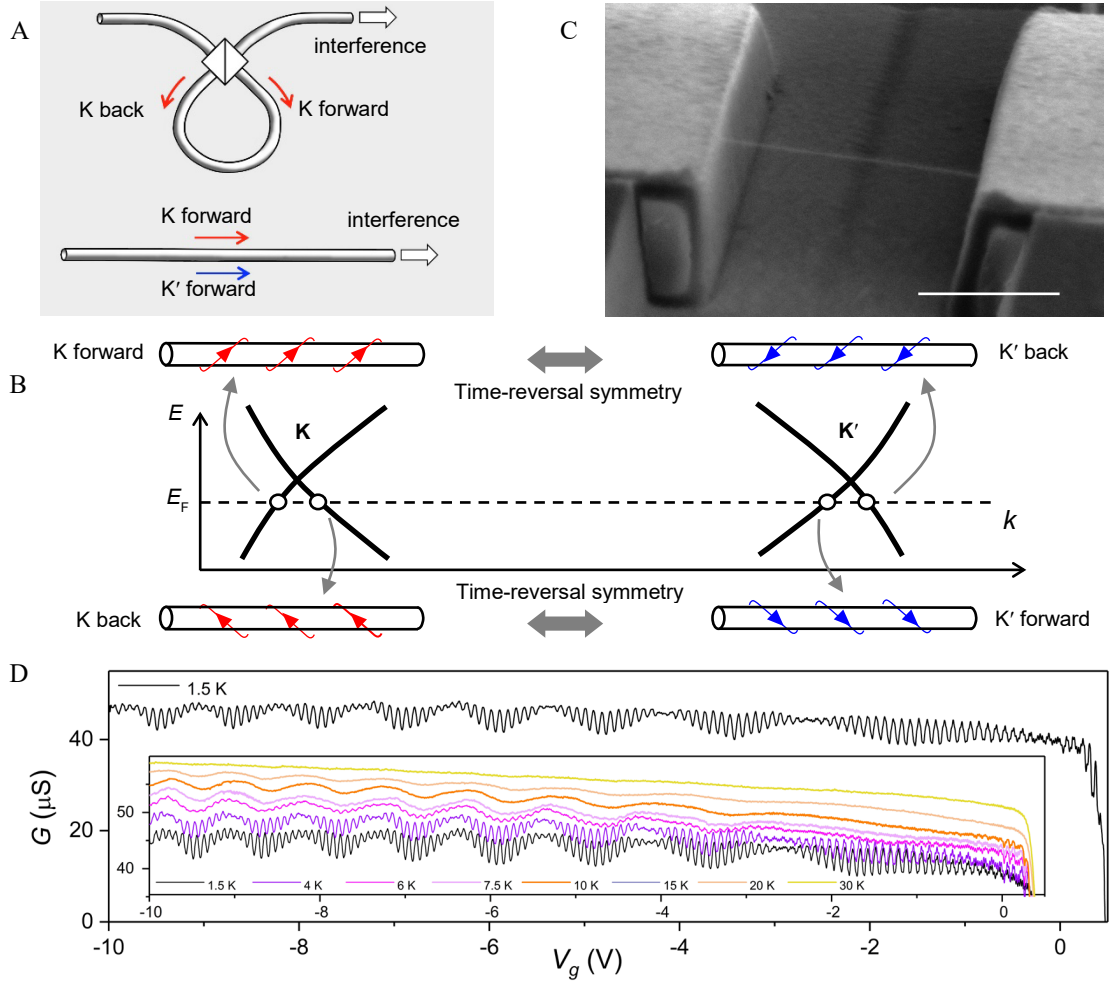

**Fig. 1. Conceptual diagram of interference mechanism and temperature dependence.**
(**A**) Geometries for realizing Sagnac interference. Top: Loop geometry realizes interference between left and right moving charge carriers within a valley. Bottom: Equivalently, a straight geometry realizes the same interference between charge carriers in the two valleys moving in the same direction. (**B**) Diagram illustrating preservation of time-reversal symmetry due to the equivalence of the forward-moving K-valley and the backward-moving K'-valley carriers. (**C**) Scanning electron microscope image of a suspended CNT, scale bar is 1 μm. (**D**) Conductance (G) versus gate-voltage ($V_g$) at T = 1.5 K for D1. Inset: Temperature dependence of fast and slow oscillations.

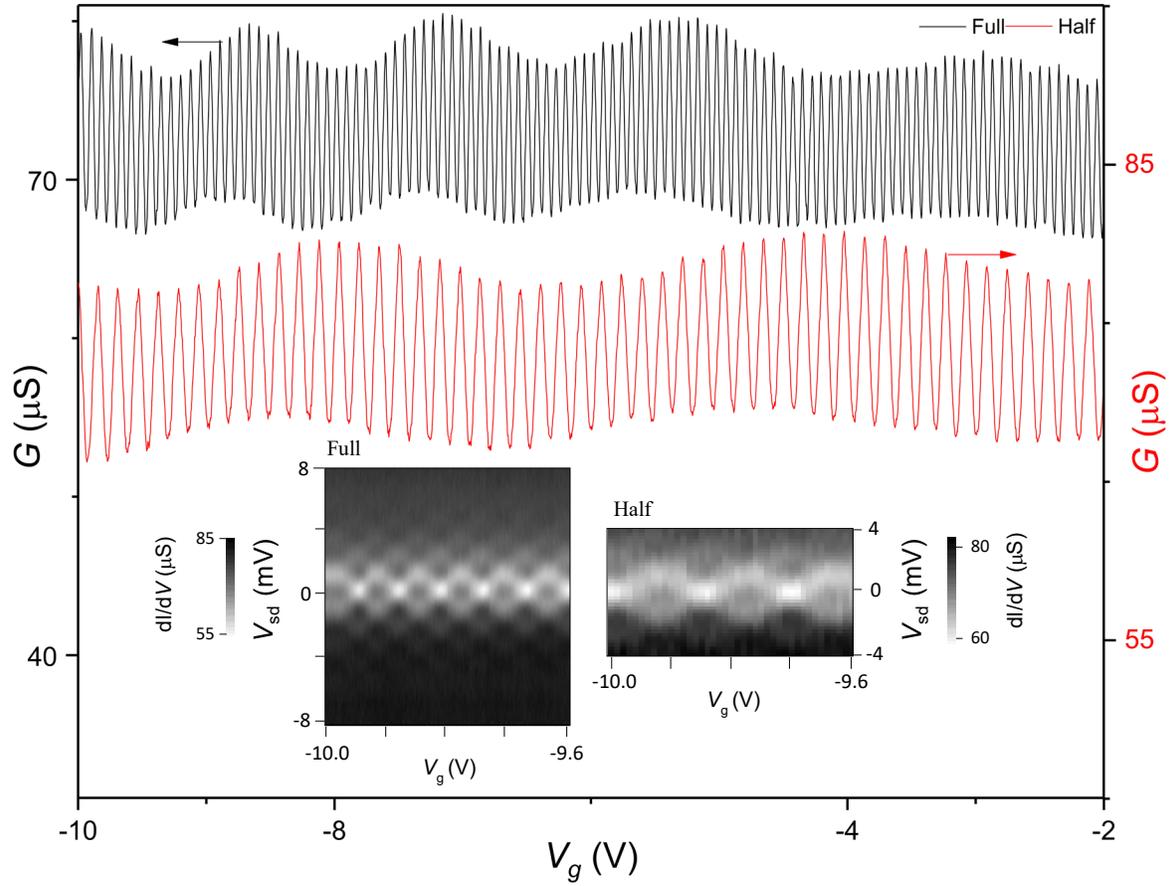

**Fig. 2. Length dependence of fast and slow oscillations.** Conductance (G) versus $V_g$ at T = 1.5 K for whole length of D2 (black) and half-length of D2 (red). Inset: Grayscale plot of differential conductance versus gate voltage $V_g$ and source-drain bias $V_{sd}$ for whole length of D2 (left) and half-length of D2 (right).

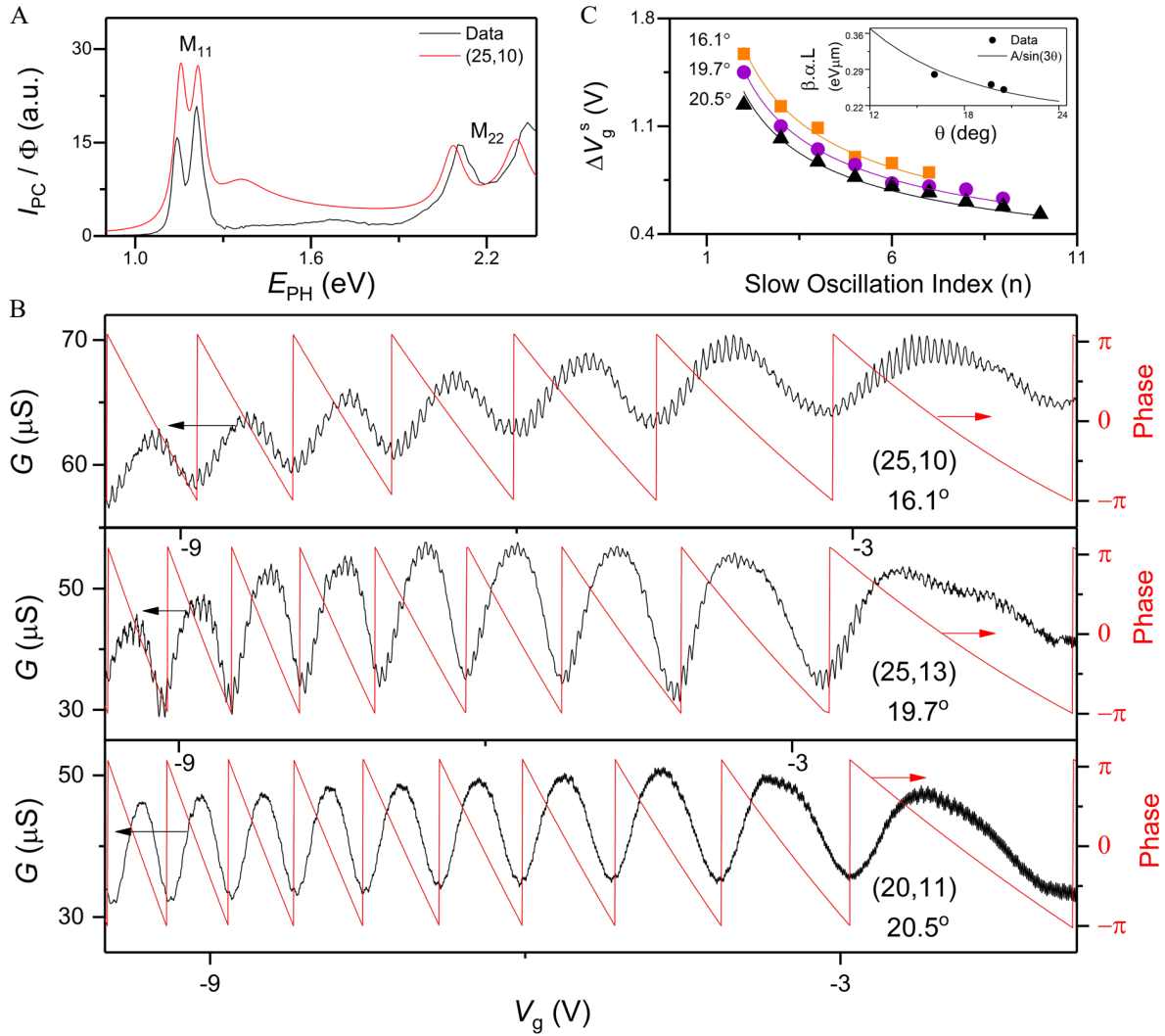

**Fig. 3. Chirality dependence and correspondence with calculated phase difference.**
(**A**) Photocurrent spectrum of D3 (25, 10) and fit to excitonic model. (**B**) Left axis: Conductance (black) in tubes D3 (25, 10), D4 (25, 13) and D5 (20,11) as a function of gate voltage $V_g$. Measured chiral angles are also noted for each device. Right axis: Calculated phase difference (red) between two interfering beams, $\Delta\varphi = k_L L - k_R L$, of the asymmetrical trigonally-warped K & K′ bands as a function of $V_g$ for the three devices. Our slow oscillations are a measure of nanotube structure in a transport measurement. (**C**) Slow oscillations period as a function of slow oscillation index for D3, D4 and D5. Fits have been obtained using $\Delta V_g^s = \beta(\sqrt{n} - \sqrt{n-1})$ where β is the fit parameter. Inset: Values of βαL for D3, D4 and D5 as a function of chiral angle θ. The solid line is a A/sin(3θ) fit versus chiral angle θ, where A is the fit parameter.

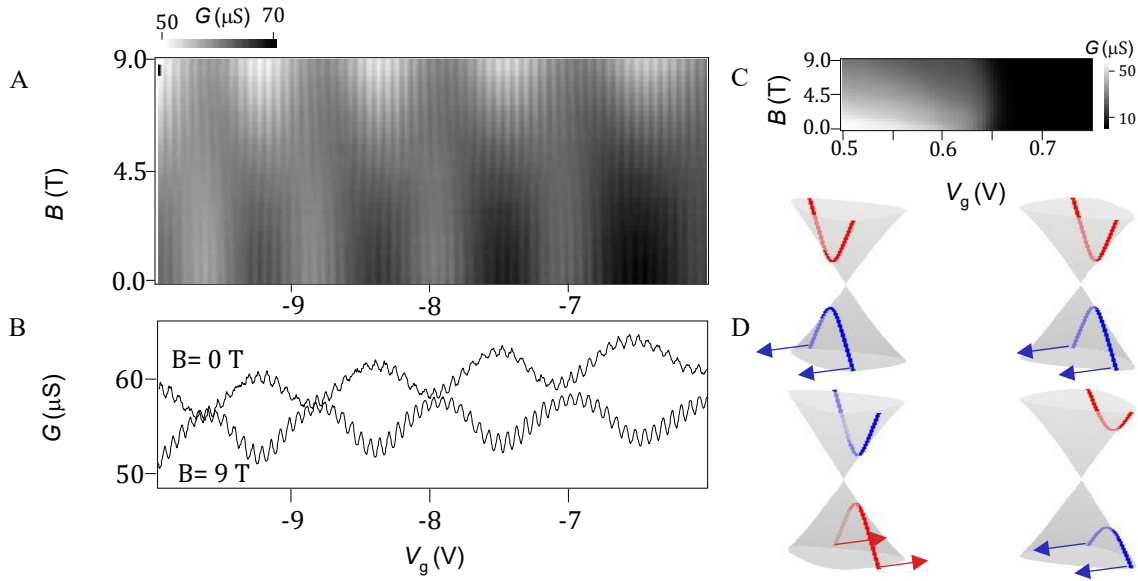

**Fig. 4. Effect of magnetic field, π phase shift in slow oscillation and proposed mechanism.** (**A**) Grayscale plot of conductance versus gate voltage $V_g$ and magnetic field B parallel to nanotube axis showing a shift in the slow oscillation at $B_{cr}$. (**B**) Line cut of (A) at B=0 T and B=9 T indicating the π phase shift. (**C**) Grayscale plot of conductance versus gate voltage $V_g$ and magnetic field B at valence band edge showing gap closing and opening at $B_{cr}$. (**D**) Top: Schematic of the energy dispersion of CNT from intersection of quantization lines with dispersion of graphene. Bottom: Quantization lines shift upon application of magnetic field parallel to the axis of the tube, resulting in changing the sign of the gap in one valley upon crossing the Dirac point of graphene. Arrows show the proposed pseudospin flip mechanism described in the text occurring even at finite doping due to the presence of a large residual many-body gap in the device at $B_{cr}$.